\documentclass[twocolumn,showpacs,preprintnumbers,amsmath,amssymb]{revtex4}

\usepackage{graphicx}
\usepackage{dcolumn}
\usepackage{bm}

\newcommand{\ket}[1]{\ensuremath{| #1 \rangle}}

\begin{document}

\title{The nuclear force imprints revealed on the elastic scattering of protons with $^{10}$C} 

\author { A. Kumar,$^1$ R. Kanungo,$^1$ A. Calci,$^2$ P. Navr\'{a}til,$^2$ A. Sanetullaev,$^{1,2}$ M. Alcorta,$^2$ V. Bildstein,$^3$ G. Christian,$^2$ B. Davids,$^2$ J. Dohet-Eraly,$^{2,4}$  J. Fallis,$^2$  A. T. Gallant,$^2$ G. Hackman,$^2$ B. Hadinia,$^3$ G. Hupin,$^{5,6}$ S. Ishimoto,$^7$ R. Kr\"{u}cken,$^{2,8}$ A. T. Laffoley,$^3$ J. Lighthall,$^2$ D. Miller,$^2$ S. Quaglioni,$^9$ J.S. Randhawa,$^1$ E. T. Rand,$^3$ A. Rojas,$^2$ R. Roth,$^{10}$ A. Shotter,$^{11}$ J. Tanaka,$^{12}$ I. Tanihata,$^{12,13}$ C. Unsworth$^2$}

\affiliation{$^1$Astronomy and Physics Department, Saint Mary's University, Halifax, NS B3H 3C3, Canada}
\affiliation{$^2$TRIUMF, Vancouver, BC V6T2A3, Canada}
\affiliation{$^3$Department of Physics, University of Guelph, Guelph, ON N1G 2W1, Canada}
\affiliation{$^4$Istituto Nazionale di Fisica Nucleare, Sezione di Pisa, Largo B. Pontecorvo 3, I-56127 Pisa, Italy}
\affiliation{$^5$Institut de Physique Nucl\'{e}aire, Universit\'{e} Paris-Sud, IN2P3/CNRS, F-91406 Orsay Cedex, France}
\affiliation{$^6$6CEA, DAM, DIF, F-91297 Arpajon, France}
\affiliation{$^7$High Energy Accelerator Research Organization (KEK), Ibaraki 305-0801, Japan}
\affiliation{$^8$Department of Physics and Astronomy, University of British Columbia, Vancouver, BC V6T 1Z1, Canada}
\affiliation{$^{9}$Lawrence Livermore National Laboratory, P.O. Box 808, L-414, Livermore, California 94551, USA}
\affiliation{$^{10}$Institut fur Kernphysik, Technische Universitat Darmstadt, 64289 Darmstadt, Germany}
\affiliation{$^{11}$University of Edinburgh, Edinburgh, United Kingdom}
\affiliation{$^{12}$RCNP, Osaka University, Mihogaoka, Ibaraki, Osaka 567 0047, Japan}
\affiliation{$^{13}$School of Physics and Nuclear Energy Engineering and IRCNPC, Beihang University, Beijing 100191, China}

\date{\today}

\begin{abstract}

How does nature hold together protons and neutrons to form the wide variety of complex nuclei in the universe? 
Describing many-nucleon systems from the fundamental theory of quantum chromodynamics has been the greatest challenge in answering this question.  The chiral effective field theory description of the nuclear force now makes this possible but requires certain parameters that are not uniquely determined. Defining the nuclear force needs identification of observables sensitive to the different parametrizations. From a measurement of proton elastic scattering on $^{10}$C at TRIUMF and {\it ab initio} nuclear reaction calculations we show that the shape and magnitude of the measured differential cross section is strongly sensitive to the nuclear force prescription. 

\end{abstract}

\maketitle

Understanding the strong nuclear force is of fundamental importance to decipher nature's way of building visible matter in our universe. 
Yet, more than a century after the discovery of the nucleus, our knowledge of the nuclear force is still incomplete. 
The formulation by Weinberg of chiral effective field theory (EFT) \cite{We91} enabled a major breakthrough in arriving at a fundamental understanding of the low-energy nuclear interactions of protons and neutrons, by forging the missing link with quantum chromodynamics. However, the question of how to best implement the theory and constrain it with experimental data remains an active topic of research, and has already led to several parameterizations of the nuclear force \cite{En03,Ma11,Ro12,Ek15,Ep15}. It is therefore important to identify experimental observables that are sensitive to different parameterizations of the chiral forces in order to reach a definitive description of the nuclear force. The study of many-nucleon systems enables a more complete understanding of the nuclear force. In particular, proton-rich and neutron-rich nuclei located at the edges of nuclear stability (drip-lines), 
can amplify less-constrained features of the nuclear force, such as its dependence on the proton-neutron asymmetry. However, there is a lack of experimental data on the properties of these systems.

Among the properties of the drip-line nuclei, 
we hypothesize in this work that the nucleon-nucleus scattering differential cross section is highly sensitive to the details of the nuclear force and hence can be used for constraining it. Indeed, it should both reveal the spectroscopic properties of the reacting system, 
such as phase shifts and their interference, as well as the effect of exotic nucleon distributions. This confluence brings a greater selectivity in the elastic scattering differential cross section
than is possible by independently investigating resonance energies, binding energies or radii. The observations reported here show that the shape and magnitude of the elastic scattering angular distribution places stringent constraints on the chiral interactions, while a study of resonance energies alone could lead to incomplete and/or misleading conclusions.  The study of elastic scattering for drip-line nuclei is however challenging because of the low-beam intensities and formulation of the {\it ab initio} structure and reaction theory. 

We report the first investigation probing the nuclear force through proton elastic scattering from $^{10}$C, located at the proton drip-line. This is an ideal system to test the effect of the nuclear force. This is because firstly, the very existence of bound $^{10}$C whose isotonic neighbours $^9$B, $^8$Be and $^{11}$N are unbound,
is a testament of the complicated strong interaction. Secondly, {\it ab initio} Green's function Monte Carlo \cite{Pi01} and no-core shell model (NCSM) \cite{Na09,Ba13} calculations have shown the three-nucleon force to be important for explaining the structure of mass number $A{=}10$ nuclei. Recent advances in {\it ab initio} nuclear reaction theory now allow us to compute the $^{10}$C(p,p) scattering cross section based on chiral forces.  Thirdly, with the low-energy re-accelerated beam available at TRIUMF, our investigation was carried out at low center-of-mass energies of $\sim$4.1 - 4.4 MeV for p+$^{10}$C, since here the low level-density of the composite (unbound) nucleus, $^{11}$N, minimizes the number of phase shifts influencing the diffraction pattern, and hence facilitates the identification of nuclear force effects, than is possible transparently with stable nuclei. Furthermore, no transfer reaction channels are open at low-energy  for this system thereby simplifying the {\it ab initio} reaction calculation. 

The experiment was performed in inverse kinematics at the ISAC rare isotope beam facility at TRIUMF \cite{Di14,Ba16} by bombarding a proton target with a $^{10}$C beam. The beam, re-accelerated using the ISAC-II superconducting linear accelerator \cite{Di14,La03}, with an average intensity of 2000 particles per second impinged on a solid hydrogen target at the IRIS reaction spectroscopy station \cite{Ka14}. A schematic of the setup is shown in Fig. 1. Energy-loss measured in a low-pressure ionization chamber allowed for clean identification of $^{10}$C from the $^{10}$B contaminant.  The beam energies at mid-target were 4.54$A$ MeV and 4.82$A$ MeV corresponding to p+$^{10}$C center of mass energies of E$_{cm}$=4.15 MeV and 4.4 MeV, respectively.  
These energies were chosen to be around the location of the 5/2$^+$ and 3/2$^-$ resonances in the $^{11}$N compound system (=$^{10}$C+p) because preliminary calculations suggested that variation of the nuclear force alters the $D_{5/2}$ and $P_{3/2}$ phase shifts and hence the cross sections, significantly. Our selected energies were chosen to be below and above the 3/2$^-$ resonance which is placed at 4.35(3) MeV in the evaluation in Ref.\cite{Ke12}. We note here however, that conflicting experimental data exist on this resonance position. Ref.\cite{Gu03} places the 3/2$^-$ resonance at 4.56(1) MeV, that is higher than both the beam energies, in which case the cross sections at the two measured energies may be similar.
 
\begin{figure}
\includegraphics[width=9cm, height=4cm]{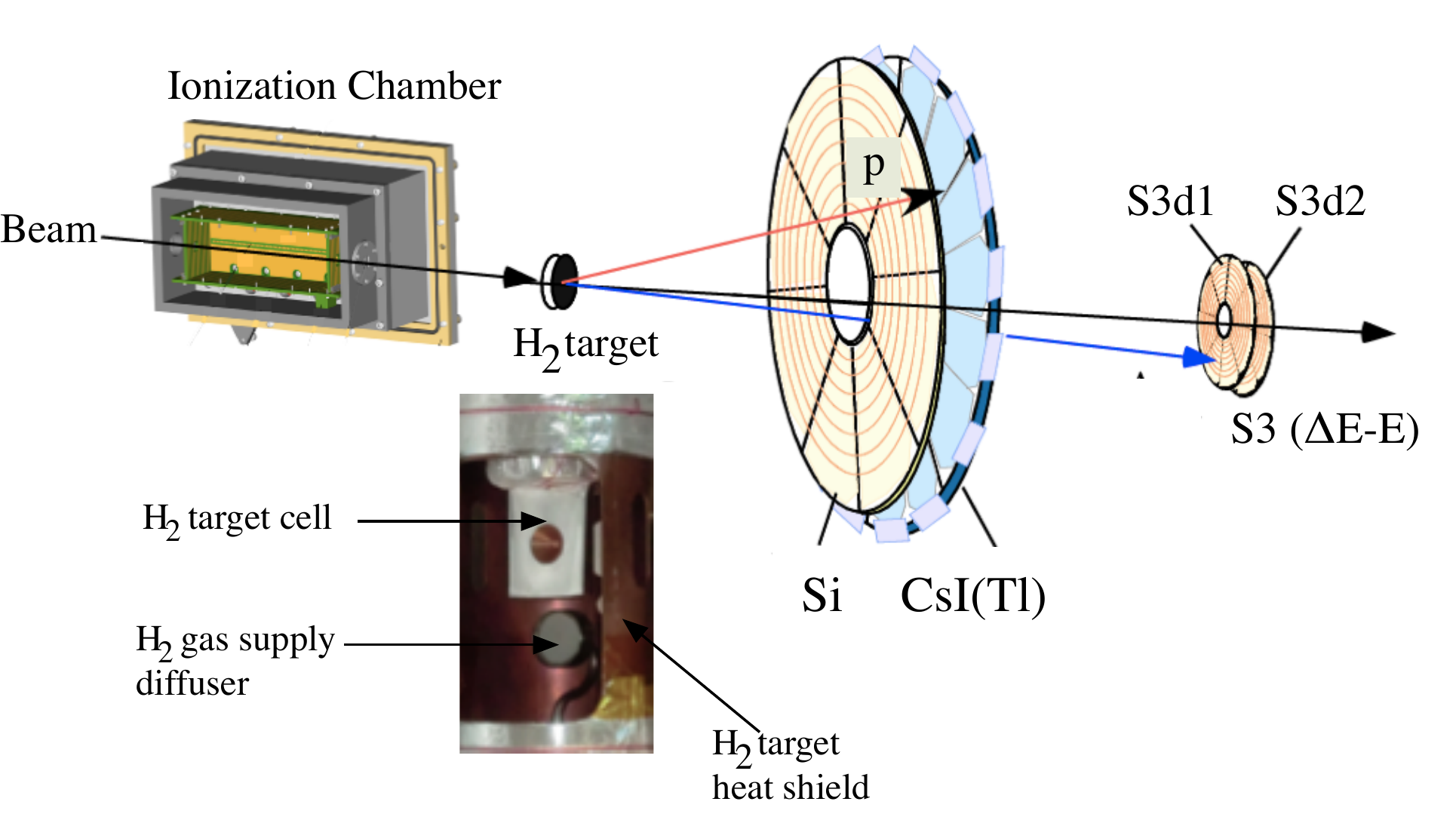}
\caption{\label{fig:epsart}  Schematic view of the experiment setup at the IRIS reaction spectroscopy station. }
\end{figure}

The scattered protons were identified using the correlation between energy-loss in an annular array of segmented silicon detectors and remaining energy deposited in CsI(Tl) detectors covering angles $\theta_{lab}$ $\sim$ 26$^\circ$ - 52$^\circ$.
The selected proton events show a very clear locus of elastic scattering (Fig. 2a). The inelastic scattering locus is only slightly visible around $\theta_{lab}$ $\sim$ 26$^\circ$ - 28$^\circ$ as most of this channel occurs at smaller $\theta_{lab}$ and was hence outside the detector coverage.  The excitation energy spectrum of $^{10}$C (Fig. 2b) was reconstructed from the measured energies and scattering angles of the protons using the missing mass technique. A small background, seen under the elastic peak, estimated by a linear fit to be $\sim$ 1 - 3 \% was subtracted to obtain the elastic scattering cross sections at the different scattering angles.  

\begin{figure}
\includegraphics[width=6cm, height=8cm]{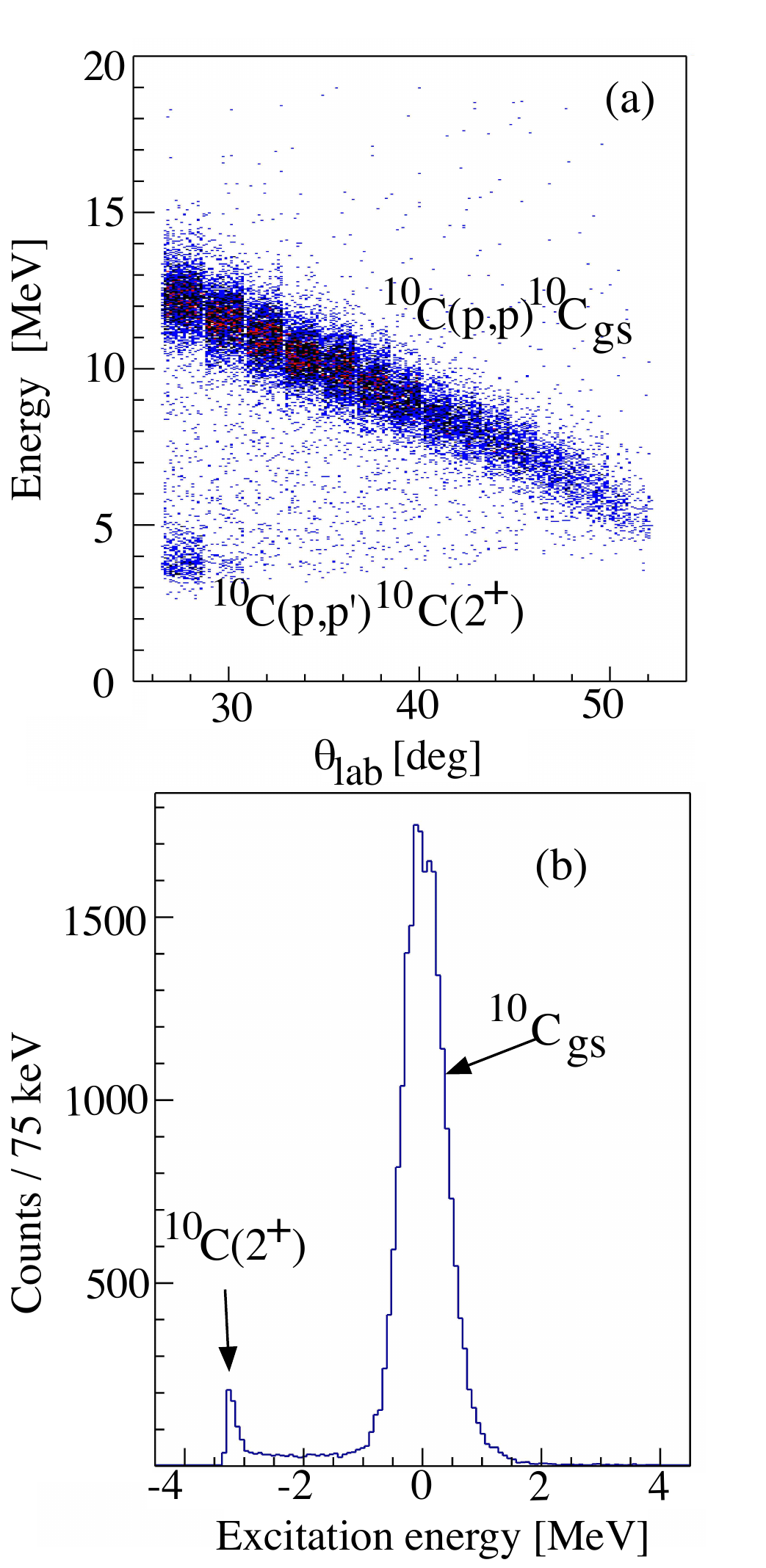}
\caption{\label{fig:epsart}  (a) The measured kinematic loci, proton energy as a function of proton scattering angle,  for $^{10}$C(p,p)$^{10}$C$_{gs}$ at E$_{cm}$ = 4.15 MeV. (b) Measured excitation energy spectrum of $^{10}$C. }
\end{figure}

The scattered $^{10}$C 
was detected by double-sided segmented silicon strip detectors (S3). The solid H$_2$ (proton) target was formed on a 5.4 $\mu$m Ag foil backing. The energy of the $^{10}$C + Ag elastic scattering peak, measured with and without H$_2$, was used to determine the H$_2$ target thickness from the energy-loss of scattered $^{10}$C through H$_2$. The  $^{10}$C + Ag scattering with presence of H$_2$ was measured simultaneously with the $^{10}$C(p,p) reaction continuously throughout the experiment, hence the target thickness at each instant was accurately known. The average target thickness was $\sim$ 80 $\mu$m. The number of incident beam particles was counted using the ionization chamber. Since the beam intensity and target thickness were measured continuously during the experiment, the absolute magnitude of the cross section is well determined. See Supplemental Material for experiment details \cite{suppl}.

The measured differential cross section for $^{10}$C(p,p) in the center-of-mass frame is shown in Fig. 3. The experimental data contain both statistical and systematic uncertainties. The systematic uncertainties are as follows : 5\% from the target thickness, 5\% from determination of the detection efficiency, and  4\% from the beam contamination, which were added in quadrature. The cross sections have similar shape and magnitude at the two different beam energies. {\it Ab initio} reaction theory calculations with three different choices of the nuclear force are shown by the curves. 

Our theoretical description of the $^{10}$C(p,p) scattering is based on the {\it ab initio} no-core shell model with continuum (NCSMC) \cite{Bar13,Bar13a,Na15}. This approach 
describes the reacting system using a basis expansion with two key components: one describing all nucleons close together, forming the $^{11}$N nucleus, and a second one describing the separated proton and $^{10}$C clusters. The former part utilizes a square-integrable basis expansion treating all 11 nucleons on the same footing. The latter part factorizes the wave function into products of $^{10}$C and proton components and their relative motion with proper scattering boundary conditions. The chiral two-nucleon (NN) and three-nucleon (3N) forces served as input for the NCSMC calculations. See Supplemental Material for more details on the calculation \cite{suppl}. 
 
In chiral EFT, the dynamics due to unresolved physics, i.e., degrees of freedom other than nucleons and pions is accounted for by contact interactions with parameters calculable in principle from QCD although presently fitted to experimental data. In particular the NN interaction is tuned to nucleon-nucleon phase shifts and the deuteron 
properties. Traditionally, the fit of the NN parameters was performed first \cite{En03}, and the 3N parameters were adjusted to $^3$H/$^3$He \cite{Ga09,Ka12} and sometimes also $^4$He data \cite{Ro12} in a second step. The NN+3N400 force (NN from Ref.\cite{En03} and 3N from Ref.\cite{Ro12}) pertains to this family of interactions and describes well the binding energy of the O, N, and F isotopes \cite{He13,Ci13}. Recently, a simultaneous NN+3N fit has been performed using not just the two-nucleon and A=3,4 data but also binding energies of $^{14}$C and $^{16,22,24,25}$O as well as charge radii of $^{14}$C and $^{16}$O \cite{Ek15}. The resulting interaction, named N$^2$LO$_{\rm sat}$, successfully describes the saturation of infinite nuclear matter \cite{Ek15}, the proton radius of the stable nucleus $^{48}$Ca \cite{Ha16} and the nuclear 
radii of neutron-rich carbon isotopes \cite{Ka16}.

We test these two parameterizations of chiral NN+3N forces and, further, investigate the impact of the chiral 3N force by comparing them with a chiral NN interaction alone. 
Fig. 3 shows that the shape and magnitude of the angular distribution is strongly influenced by the nuclear force prescription. The results obtained with the NN interaction from Ref. \cite{En03} (black dotted curves) show a strong dip in the cross section at $\theta_{cm}$ $\sim$ 80$^{\circ}$, while no such feature is observed in the data. The addition of the 3N force with a momentum cut-off of 400 MeV [6] (NN+3N400, blue long dashed curves) produces a much different shape. This shows the strong influence of the three-nucleon interaction on the angular distribution. 
While it improves the overall agreement with the data the addition of the 3N force in the NN+3N400 interaction clearly still does not explain the observed angular distribution characteristics. Meanwhile with the N$^2$LO$_{\rm sat}$ NN+3N interaction (red solid curves) 
the predicted shape of the angular distribution is in very good agreement with the data as evidenced by the scaled result (red dashed curves). The magnitude of the cross section is higher than the data which, as discussed below, reflects effects of $S_{1/2}$ phase shift(s) and hence places further constraint on the force prescription.

\begin{figure}
\includegraphics[width=6cm, height=8cm]{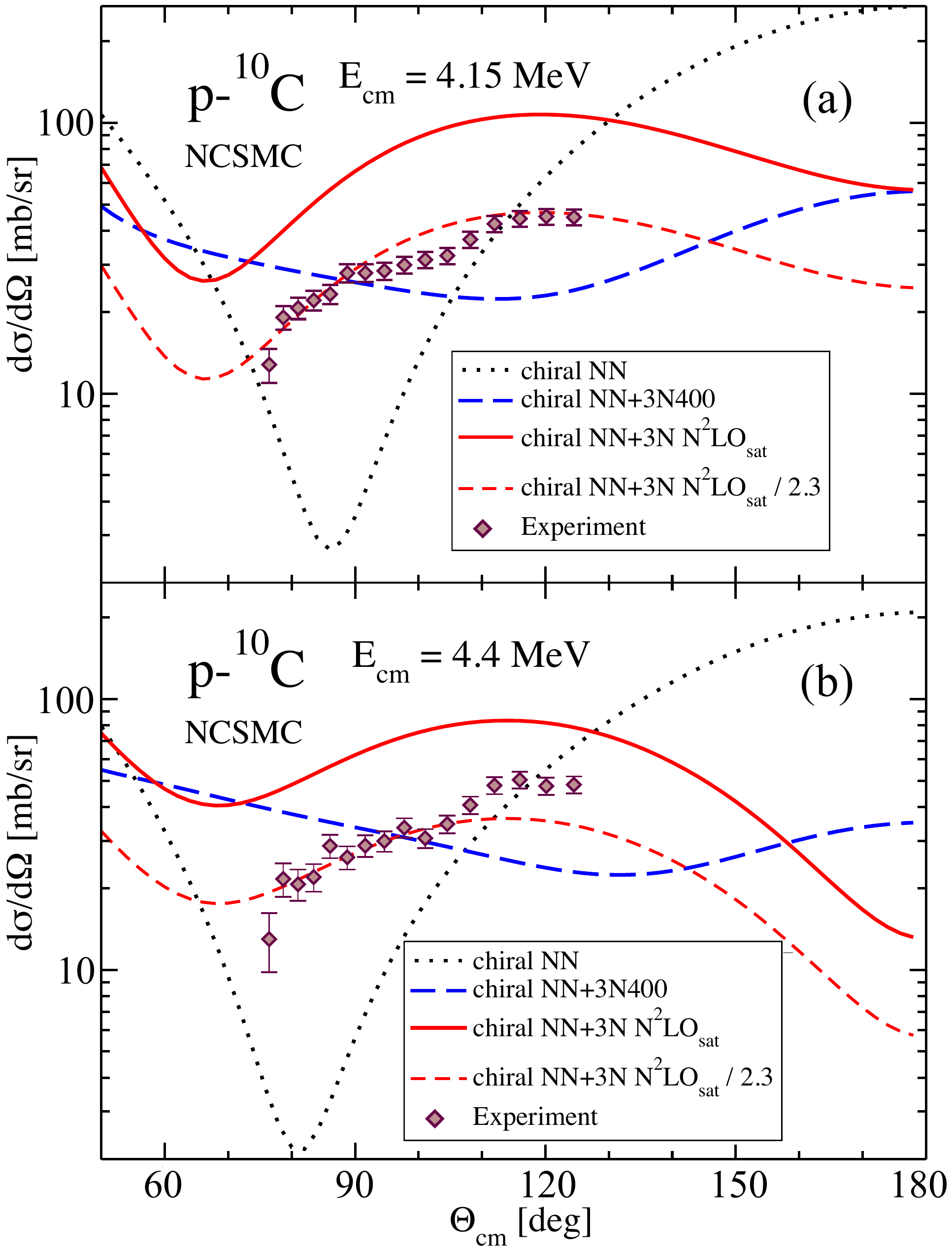}
\caption{\label{fig:3epsart}  (a) Measured differential cross section for $^{10}$C(p,p)$^{10}$C$_{gs}$ at (a) E$_{cm}$ = 4.15 and (b) E$_{cm}$= 4.4 MeV. The curves are {\it ab initio} theory calculations. The black dotted / blue long dashed / red solid curves are with the chiral NN / NN+3N400 / N$^2$LO$_{\rm sat}$ interactions. The red dashed curve is the N$^2$LO$_{\rm sat}$ calculation scaled down by a factor of 2.3.}
\end{figure}

Compared to calculations with the NN interaction, the N$^2$LO$_{\rm sat}$ and NN+3N400 forces result in only a small $\sim$ 1\% effect in static properties, such as the proton and matter radii of $^{10}$C. While the binding energy predicted by NN+3N400 is lower by $\sim$ 12\% than the predictions from N$^2$LO$_{\rm sat}$, the latter is $\sim$ 3\% higher than the experimental value, thus a binding energy-based distinction of the different forces is not straightforward. However, here we find a strong constraint on the nuclear force emerging from the dramatic change of the angular distribution shape (Fig. 3). The chi-square ($\chi^2$) values of scaled cross sections show this clearly. For E$_{cm}$=4.15 MeV, the best-fit $\chi^2$ = 65.5 for the NN, 25.9 for the NN+3NF400 and 1.7 for the N$^2$LO$_{\rm sat}$ interactions (after its scaling), making N$^2$LO$_{\rm sat}$ the best selection among the three.  Additionally, the failure of the NN+3N400 interaction to reproduce the angular distribution despite good predictions of the binding energies of the oxygen isotopes \cite{He13,Ci13}, shows the limited selectivity of binding energies in differentiating among the nuclear force models. The magnitude of the cross section however shows the deficiencies of the N$^2$LO$_{\rm sat}$ interaction.

Calculated phase shifts with the three interactions
are shown in Fig. 4. The resonance energies from R-matrix analysis are listed in Table 1. 
In the energy region of the experiment (vertical lines in Fig. 4), the shape of the angular distribution is dominated by the $3/2^-$ and the $5/2^+$ phase shifts. The $5/2^+$ resonance couples strongly the $^2D_{5/2}$($^{10}$C($0^+$)) and $^6S_{5/2}$($^{10}$C($2^+_1$)) partial waves. Therefore we present the $5/2^+$ eigenphase shifts. The other shown resonances are dominated by a single (shown) partial wave with the exception of the N$^2$LO$_{\rm sat}$ $3/2^-$ that couples $^2P_{3/2}$($^{10}$C($0^+$)) and $^6P_{3/2}$($^{10}$C($2^+_1$)) partial waves. 

\begin{figure}
\includegraphics[width=6cm, height=8cm]{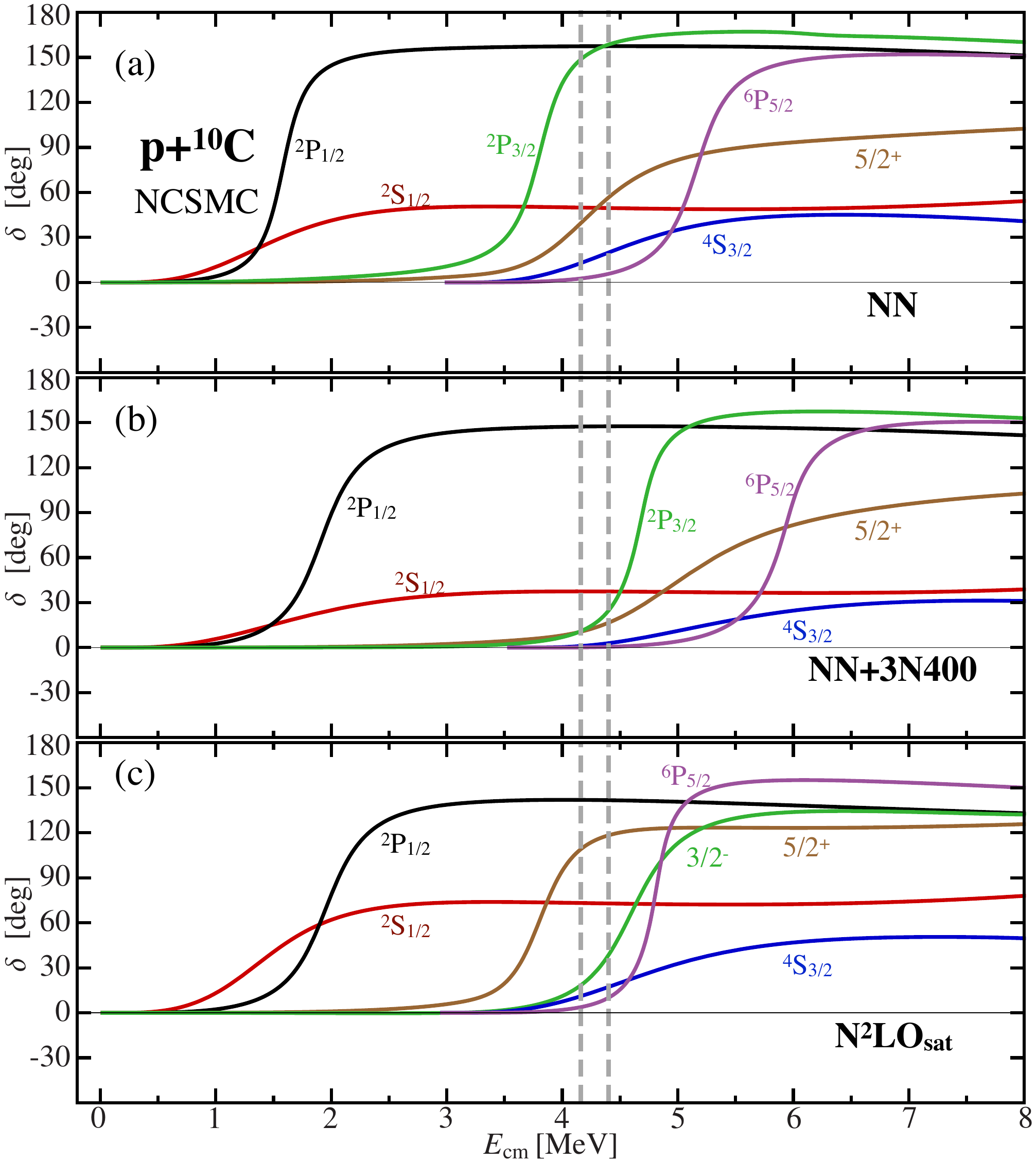}
\caption{\label{fig:4epsart} Calculated p+$^{10}$C phase shifts (eigenphase shifts for $5/2^+$ and N$^2$LO$_{\rm sat}$ $3/2^-$). Ab initio NCSMC results obtained using chiral NN (panel a), chiral NN+3N400 (panel b) and chiral N$^2$LO$_{\rm sat}$ (panel c) are compared. The vertical dashed lines show the energies where the experiment was performed.} 
\end{figure}

We observe that the $5/2^+$ and $3/2^-$ resonances are placed differently in the three calculations. Using the chiral NN interaction alone (Fig. 4(a)), the $3/2^-$ resonance is below the $5/2^+$ one and below the $^{10}$C(p,p) experimental region. Switching on the chiral 3N force in the NN+3N400 calculation  (Fig. 4(b)), the two resonances are almost degenerate and slightly above the region of measurement. With the chiral N$^2$LO$_{\rm sat}$ (NN+3N) interaction, the $5/2^+$ resonance is below the $3/2^-$ one and below the energy region where 
the $^{10}$C(p,p) measurements were performed (Fig. 4(c)). 
Only using the N$^2$LO$_{\rm sat}$ interaction the ordering of the $5/2^+$ and the $3/2^-$ is in qualitative agreement with the established ordering of the isospin analog resonances in the mirror $^{11}$Be nucleus~\cite{Ke12,Ca16}. 

A comparison of the computed $3/2^-$ and $5/2^+$ resonance properties (Table 1) to the evaluated data therefore, could erroneously lead one to believe that the N$^2$LO$_{\rm sat}$ interaction works almost perfectly. However, the magnitude of our measured cross section is overestimated by the N$^2$LO$_{\rm sat}$ calculations. Hence, it should be emphasized that the present experiment tests the nuclear force more strictly than a straight comparison of energies and widths of the resonances. This is because the differential cross section receives also contributions from phase shifts in other partial waves. In this case, in particular from the $^2S_{1/2}$ that contributes only to the magnitude of the cross section and is much more pronounced in the N$^2$LO$_{\rm sat}$ calculation.

\begin{table}
\caption{\label{tab:table1} Energies ($E_r$) and widths ($\Gamma$) in MeV of low-lying resonances of $^{11}$N.}
\begin{ruledtabular}
\begin{tabular}{ccccccc}
               &  \multicolumn{2}{c}{NN+3N400}&  \multicolumn{2}{c}{N$^2$LO$_{\rm sat}$} & \multicolumn{2}{c}{Data evaluation \cite{Ke12}}  \\
 $J^\pi$    &   $E_r$     &    $\Gamma$      & $E_r$     &    $\Gamma$      & $E_r$      &  $\Gamma$   \\
\hline
$1/2^+$ &  1.29 & 2.85  &   1.33               &    1.45                          & 1.49(6)               &  0.83(3)                  \\
 $1/2^-$ &   1.91 & 0.54  &  1.95	             &    0.57	                    & 2.22(3)	         &  0.6(1)     \\
 $5/2^+$ &   4.89& 1.76   &3.81	             &    0.53	                    &  3.69(3)	         &  0.54(4)    \\
 $3/2^-$	&   4.62& 0.47 &4.60	             &    0.70	                    &  4.35(3)	         &  0.34(4)    \\
 $3/2^+$ &   5.88& 4.09 &4.39	             &    2.55	                    &  N/A	                 &  N/A         \\
 $5/2^-$	&   5.85& 0.66&4.77	             &    0.41	 	            &                           &                 \\
\end{tabular}
\end{ruledtabular}
\end{table}

The present reaction calculation does not include the $^9$Be+2p breakup channel, which lies just a few hundred keV below the energy of the experiment. This omission contributes only a small part to the over-prediction of the data by the N$^2$LO$_{\rm sat}$ interaction. An estimate of this is obtained from, our calculated $^{10}$C(p,p')$^{10}$C(2$^+$) inelastic cross section and is only a few mb/sr at the same relative energy.  Therefore, given the reasonable convergence of our calculations, this shows
that the N$^2$LO$_{\rm sat}$ interaction, though it provides the best fit of the present data angular distribution shape, is still missing a complete description of the nuclear force. This deficiency becomes apparent with this angular distribution data and is not possible to judge based on resonance energies alone.

We should, however, make it clear that the N$^2$LO$_{\rm sat}$ interaction indeed captures some important missing physics compared to the other chiral interaction (NN+3N400). It provides a more realistic description of the nuclear density and smaller gaps between major harmonic-oscillator shells.
Overall, we observe that none of the available parameterizations of the chiral nuclear force is optimal in all aspects. There is a significant progress in the development of high-quality chiral NN potentials; the N$^4$LO order has now been reached \cite{Ep15,En15}. These potentials achieve an excellent description of the NN system. However, despite this progress \cite{Kr12,Bi16}, a chiral 3N force parameterization matching their quality is still missing. 

In summary, with the measured angular distribution of low-energy elastic scattering off extremely exotic $^{10}$C nuclei we have demonstrated for the first time a strong sensitivity of this scattering to the nuclear force prescription. The low-level density and neutron-proton asymmetry in drip-line nuclei like $^{10}$C bring in new and greater sensitivity to the nuclear force, allowing for discriminating between the different chiral interactions and finding further constraints for them. The measured $^{10}$C(p,p)$^{10}$C$_{gs}$ differential cross section shows that only the N$^2$LO$_{\rm sat}$ interaction provides an angular distribution shape consistent with the experiment but fails to reproduce its magnitude. This suggests that N$^2$LO$_{\rm sat}$ is improved compared to the other forces but is still not an adequate description of the nuclear force.
The new finding of this large sensitivity of the angular distribution will trigger more intensive efforts in {\it ab initio} calculations to single-out which parameters and components of the chiral interactions are responsible for the successful description of the $^{10}$C(p,p) data. 
Extreme systems, such as the $^{11}$N and $^{10}$C(p,p) investigated here both experimentally and theoretically, thus provide one of the most stringent tests of the quality of the present and new generations of nuclear forces.

The authors express sincere thanks to the TRIUMF beam delivery team. The support from Canada Foundation for Innovation, NSERC, Nova Scotia Research and Innovation Trust and the DFG through SFB 1245 is gratefully acknowledged. TRIUMF receives funding via a contribution through the National Research Council Canada. Computing support came from the LLNL institutional Computing Grand Challenge Program, from an INCITE Award on the Titan supercomputer of the Oak Ridge Leadership Computing Facility (OLCF) at ORNL and from Calcul Quebec and Compute Canada. The work is prepared in part by LLNL under Contract DE-AC52-07NA27344. This material is based in part upon work supported by the U.S. Department of Energy, Office of Science, Office of Nuclear Physics, under Work Proposal  No. SCW1158.



\begin{thebibliography}{99}

\bibitem{We91} S. Weinberg, Nucl. Phys. {\bf B 363}, 3 (1991).
\bibitem{En03} D.R. Entem and R. Machleidt, Phys. Rev. {\bf C 68}, 041001 (2003).
\bibitem{Ma11} R. Machleidt and D.R. Entem,  Phys. Rep. {\bf 503}, 1 (2011).
\bibitem{Ro12} R. Roth et al., Phys. Rev. Lett. {\bf 109}, 052501 (2012).
\bibitem{Ek15} A. Ekstr\"{o}m et al., Phys. Rev. {\bf C 91}, 051301(R) (2015). 
\bibitem{Ep15} E. Epelbaum, H. Krebs and U.-G. Mei{\ss}ner, Phys. Rev. Lett. {\bf 115}, 122301 (2015).
\bibitem{Pi01} S.C. Pieper and R.B. Wiringa, Ann. Rev. Nucl. Part. Sci 51, {\bf 53} (2001).
\bibitem{Na09} P. Navr\'{a}til, S. Quaglioni, I. Stetcu and B. R. Barrett, J. Phys. G: Nucl. Part. Phys. {\bf 36}, 083101 (2009).
\bibitem{Ba13} B.R. Barrett, P. Navr\'{a}til, and J.P. Vary, Prog. Part. Nucl. Phys. {\bf 69}, 131 (2013).
\bibitem{Di14} J. Dilling and R.Krücken, Hyperfine Interac. {\bf 225}, 111 (2014).
\bibitem{Ba16} G.C. Ball, G. Hackman, and R. Kr\"{u}cken,  Phys. Scr. {\bf 91}, 09302 (2016).
\bibitem{La03} R.E. Laxdal et al.,  Nucl. Instrm. And Meth. {\bf B 204}, 400  (2003).
\bibitem{Ka14} R. Kanungo, Hyperfine Interac. {\bf 225}, 235 (2014).
\bibitem{Ke12} J.H. Kelley et al.,  Nucl. Phys. {\bf A 880}, 88 (2012).
\bibitem{Gu03} V. Guimar\~{a}es et al., Phys. Rev. {\bf C 67}, 064601 (2003).
\bibitem{Bar13} S. Baroni, P. Navr\'{a}til and S. Quaglioni,  Phys. Rev. Lett. {\bf 110}, 022505 (2013).
\bibitem{Bar13a} S. Baroni, P. Navr\'{a}til and S. Quaglioni, Phys. Rev. {\bf C 87}, 034326 (2013).
\bibitem{Na15} P. Navr\'{a}til, S. Quaglioni, G. Hupin, C. Romero-Redondo, and A. Calci, Phys. Scr. {\bf 91}, 053002 (2016).
%
\bibitem{suppl} See Supplemental Material 
for some details of the calculations, and experiment which includes Refs. [20-39]. 
\bibitem{Wi77} K. Wildermuth and Y. C. Tang, (Viewig+Teubner Verlag, Braunschweig, 1977) (ISBN 978-3-528-08373-1).
\bibitem{Ta78} Y. C. Tang, M. LeMere and D. R. Thompsom,  Phys. Rep. {\bf 47}, 167 (1978).
\bibitem{Fl82} T. Fliessbach and H. Walliser, Nucl. Phys. {\bf A 377}, 84 (1982).
\bibitem{La86} K. Langanke and H. Friedrich, (Advance in Nuclear Physics) ed J. W. Negele and E. Vogt (New York: Springer, 1986) ch 17, pp 223-363 (ISBN 978-0-306-45157-7).
\bibitem{De10}  P. Descouvemont and D. Baye, Reports on Progress in Physics {\bf 73}, 036301 (2010). 
\bibitem{He98} M. Hesse, J.-M. Sparenberg, F. Van Raemdonck, and D. Baye, Nucl. Phys. {\bf A 640}, 37 (1998).
\bibitem{He02} M. Hesse, J. Roland, and D. Baye, Nucl. Phys. {\bf A 709}, 184 (2002). 
\bibitem{Na07} P. Navr\'{a}til, Few Body Syst. 41, 117 (2007).
\bibitem{We94} F. Wegner, Ann. Phys. (Leipzig) {\bf 506}, 77 (1994).
\bibitem{Bo07} S. K. Bogner, R. J. Furnstahl, and R. J. Perry,  Phys. Rev. {\bf C 75}, 061001(R) (2007).
\bibitem{Sz00} S. Szpigel and R. J. Perry, Quantum Field Theory. A 20th Century Profile, edited by A. N. Mitra (Hindustan Publishing Co., New Delhi, 2000).
\bibitem{Hu13} G. Hupin, J. Langhammer, P. Navr\'{a}til, S. Quaglioni, A. Calci, and R. Roth,  Phys. Rev. {\bf C 88}, 054622 (2013).
\bibitem{Bi14} S. Binder, J. Langhammer, A. Calci, and R. Roth, Phys. Lett. {\bf B 736}, 119 (2014).
\bibitem{La15} J. Langhammer, P. Navr\'{a}til, S. Quaglioni, G. Hupin, A. Calci, and R. Roth,  Phys. Rev. {\bf C 91}, 021301(R) (2015).
\bibitem{Ca06} E. Casarejos et al., Phys. Rev. {\bf C 73}, 014319 (2006).
\bibitem{Ma00} K. Markenroth et al.,  Phys. Rev. {\bf C 62}, 034308 (2000).
\bibitem{Le98} A. L\'{e}pine-Szily et al., Phys. Rev. Lett. {\bf 80}, 1601 (1998).
\bibitem{Ol00} J. M. Oliveira et al., Phys. Rev. Lett. {\bf 84}, 4056 (2000).
\bibitem{Sc81} B. I. Schneider, Phys. Rev. {\bf A 24}, 1 (1981).
\bibitem{Do16} J. Dohet-Eraly , P. Navr\'{a}til, S. Quaglioni, W. Horiuchi, G. Hupin, and F. Raimondi, Phys. Lett. {\bf B 757}, 430 (2016).
%
\bibitem{Ga09} D. Gazit, S. Quaglioni, and P. Navr\'{a}til, Phys. Rev. Lett. {\bf 103}, 102502 (2009).
\bibitem{Ka12} N. Kalantar-Nayestanaki, E. Epelbaum, J. G. Messchendorp, and A. Nogga,  Rep. Prog. Phys. {\bf 75}, 016301 (2012).
\bibitem{He13} H. Herhert et al., Phys. Rev. Lett. {\bf 110}, 242501 (2013).
\bibitem{Ci13} A. Cipollone, C. Barbieri and P. Navr\'{a}til, Phys. Rev. Lett. {\bf 111}, 062501 (2013).
\bibitem{Ha16} G. Hagen et al., Nature Physics {\bf 12}, 186 (2016).
\bibitem{Ka16} R. Kanungo et al., Phys. Rev. Lett. {\bf 117}, 102501 (2016).
\bibitem{Ca16} A. Calci, P. Navr\'{a}til, R. Roth, J. Dohet-Eraly,  S. Quaglioni, and G. Hupin,  Phys. Rev. Lett. {\bf 117}, 242501 (2016).
\bibitem{En15} D. R. Entem, N. Kaiser, R. Machleidt, and Y. Nosyk,  Phys. Rev. {\bf C 91}, 014002 (2015).
\bibitem{Kr12} H. Krebs, A. Gasparyan, and E. Epelbaum,  Phys. Rev. {\bf C 85}, 054006 (2012).
\bibitem{Bi16} S. Binder et al., Phys. Rev. {\bf C 93}, 044002 (2016).



\end{thebibliography}

\clearpage

\centerline{ \Large {\bf Supplemental Material}}

\section{DETAILS OF THE CALCULATIONS.}

Our approach to the description of the p+$^{10}$C scattering is based on combining the {\it ab initio} no-core shell model (NCSM)~\cite{Na09,Ba13} and the resonating group method (RGM)~\cite{Wi77,Ta78,Fl82,La86}. The state-of-the-art version of this approach, the {\it ab initio} no-core shell model with continuum (NCSMC), has been introduced quite recently~\cite{Bar13,Bar13a,Na15}. 

The NCSMC is based on an expansion of the $A$-nucleon, here $^{11}$N, wave function consisting of a square-integrable $^{11}$N NCSM part and a cluster part, here $^{10}$C+p, with the proper asymptotic behavior. The $^{10}$C wave function is also obtained by the NCSM with the same Hamiltonian applied to the whole system. The 11-nucleon wave function is represented as the generalized cluster expansion
\begin{align}
\ket{\Psi^{J^\pi T}_{A\texttt{=}11}} = &  \sum_\lambda c^{J^\pi T}_\lambda \ket{^{11} {\rm N} \, \lambda J^\pi T} \nonumber \\
& +\sum_{\nu}\!\! \int \!\! dr \, r^2 
                 \frac{\gamma^{J^\pi T}_{\nu}(r)}{r}
                 {\mathcal{A}}_\nu \ket{\Phi^{J^\pi T}_{\nu r}} \,.\label{eq:ansatz}
\end{align}
The first term consists of an expansion over the NCSM eigenstates of the compound system ($^{11}$N) indexed by $\lambda$. These states are well suited to cover the localized correlations of the 11-body system, but are inappropriate to describe clustering and scattering properties. The latter properties are addressed by the second term corresponding to an expansion over the antisymmetrized channels with
\begin{align}
\ket{\Phi^{J^\pi T}_{\nu r}} = & 
		\Big[ \!\! \left(
        		\ket{^{10} {\rm C} \, \lambda_1 J_1^{\pi_1}T_1}\ket{p \, \tfrac12^{\texttt{+}}\tfrac12}
         	\right)^{(sT)} Y_\ell(\hat{r}_{10,1}) \Big]^{(J^{\pi}T)} \nonumber\\ 
	& \times\,\frac{\delta(r{-}r_{10,1})}{rr_{10,1}} \; ,
\label{eq:rgm-state}
\end{align}
which describe the $^{10}$C+p in the relative motion. Here, $\vec{r}_{10,1}$ is the separation between the center-of-mass of $^{10}$C and the proton and $\nu$ is a collective index for the relevant quantum numbers. The discrete expansion coefficients   and the continuous relative-motion amplitudes   are obtained as a solution of the generalized eigenvalue problem derived by representing the Schr\"{o}dinger equation in the model space of expansion (1)~\cite{Na15}. The resulting NCSMC equations are solved by the coupled-channel R-matrix method on a Lagrange mesh~\cite{De10,He98,He02}. A particular technically complex task is the inclusion of the three-nucleon (3N) interaction in the calculations. For nuclei with $A{>}5$ we rely on the formalism using uncoupled densities of Ref.~\cite{Hu13}. As described in the main section of the letter, we employ two kinds of chiral EFT Hamiltonians: the N$^3$LO NN of Ref.~\cite{En03,Ma11} combined with the N$^2$LO 3N with a 400 MeV cutoff (NN+3N400) of Refs.~\cite{Ro12,Na07,Ga09} and the recently developed N$^2$LO$_{\rm sat}$ NN+3N interaction of Ref.~\cite{Ek15}. To test the impact of the chiral 3N interaction, we also perform calculations using just the chiral N3LO NN. We apply the similarity renormalization group (SRG) technique~\cite{We94,Bo07,Sz00} to soften the chiral interaction and keep two- and three-body SRG induced terms in all calculations, even in the case when the initial chiral 3N is switched off. 

\begin{figure}
\includegraphics[width=6cm, height=4cm]{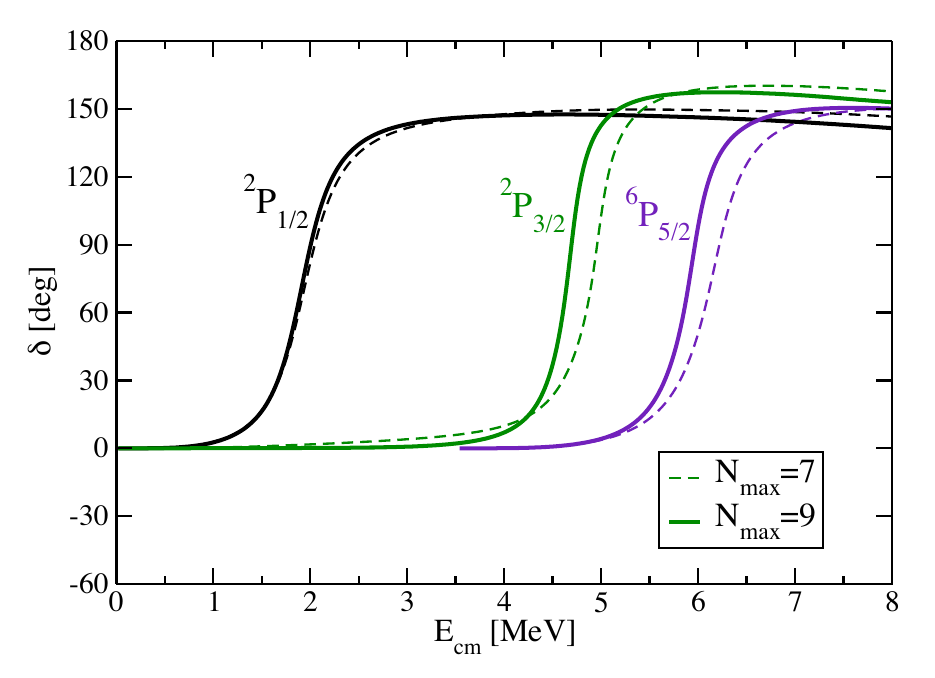}
\caption{\label{fig:suppl1} Calculated p+$^{10}$C phase-shift dependence on the HO basis expansion parameter $N_{\rm max}$ using the NN+3N400 interaction. The $N_{\rm max}{=}9$ basis is the largest currently technically feasible within the NCSMC.}
\end{figure}

The present NCSMC calculations are performed including the first three NCSM eigenstates ($0^+$, $2^+_1$ , $2^+_2$) of the $^{10}$C nucleus, entering the cluster states in Eq~ (\ref{eq:rgm-state}) and at least the first six negative- and three positive-parity NCSM eigenstates of $^{11}$N (entering the first term on the right-hand-side of Eq.~(\ref{eq:ansatz})). We select the parameters of the calculations so that the SRG unitarity is (almost) preserved and the basis-size convergence is (almost) reached. In particular, we use the harmonic-oscillator (HO) frequency of $\hbar\Omega{=}20$~MeV, the SRG evolution parameter $\Lambda{=}2$~fm$^{-1}$, and the $N_{\rm max}$ in the range from 4 to 9. This parameter choice was established in numerous studies for various nuclei~\cite{Hu13,Bi14,La15}. See Fig.~\ref{fig:suppl1} for the phase-shift convergence demonstration.

Experimentally, the resonances in $^{11}$N have been observed through the resonant scattering of p+$^{10}$C~\cite{Ca06,Ma00} and transfer reactions~\cite{Gu03,Le98,Ol00}. The observed (unbound) ground state energy ranges from 1.3 MeV to 1.6 MeV between the various experiments with the first excited state being around 2.2-2.5 MeV. The $5/2^+$ resonance is observed at 3.69(3) MeV and the $3/2^-$ resonance is found at 4.35(3) MeV~\cite{Ke12} that are closer to the predictions with the N$^2$LO$_{\rm sat}$ interaction. We note that the similarity of the shapes of the angular distribution data at the two measured energies in this experiment places a constraint on the lower limit of the resonance energy for the $3/2^-$ level, and indicates that the resonance energy is likely above 4.4 MeV, more in line with the determination from Ref.~\cite{Gu03}, of 4.56(1) MeV. The resonance energies and widths corresponding to the NN+3N400 and N$^2$LO$_{\rm sat}$ calculations (Fig.~4(b) and 4(c), respectively in the main article) are compared to the evaluated ones~\cite{Ke12} in Table 1 in the main article. From Fig.~\ref{fig:suppl1}, we infer the uncertainty on the energy and the width of resonances in the region of interest to be $\Delta E_r{\sim}250$~keV and $\Delta\Gamma{\sim}50$~keV. Such uncertainties are much smaller than the difference between the $5/2^+$ and the $3/2^-$ resonance centroids in the N$^2$LO${\rm sat}$ calculation, as seen from Table 1 in the main article. This supports the robustness of the shapes of the calculated cross sections, which are determined mostly by the relative position of the two resonances. 

The NCSMC calculated values were obtained by S-matrix analysis in the complex plane~\cite{Sc81,Do16}. The assignment of $5/2^-$ is uncertain~\cite{Gu03,Ol00} and no information is available for $3/2^+$ in the evaluation.

\section{DETAILS OF EXPERIMENT.}

\subsection{Solid hydrogen target}
The solid H$_2$ reaction target was formed by freezing H$_2$ gas onto a 5.4 $\mu$m Ag foil. The copper target cell was cooled to $\sim$4K before depositing H$_2$. The cell is surrounded by a copper heat shield, maintained at a temperature of $\sim$30 K, to reduce radiative heating from the ambient room temperature. The heat shield masks a few parts of the silicon detector array that detects the scattered protons. The geometric efficiency of the detector was therefore determined using Monte Carlo simulation of the experiment setup.

\subsection{Beam Identification.}

\begin{figure}
\includegraphics[width=6cm, height=6cm]{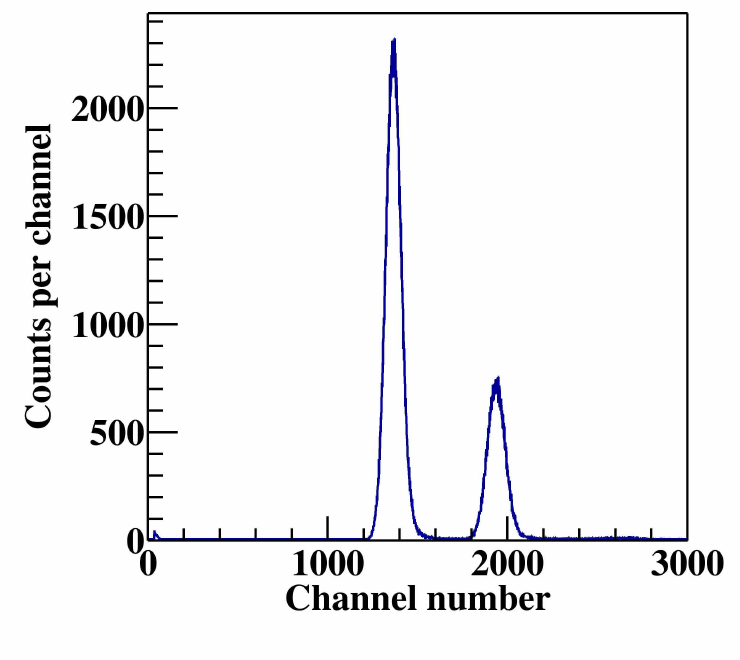}
\caption{\label{fig:suppl2} Spectrum of energy loss measured using the ionization chamber shown in arbitrary units. Left peak is $^{10}$B beam and right peak in $^{10}$C beam.}
\end{figure}

The beam was a cocktail of $^{10}$C with $^{10}$B as contaminant. This isobaric contaminant was identified using the energy-loss measured with a low-pressure ionization chamber filled with isobutene gas. 
Figure~\ref{fig:suppl2} shows the measured energy loss spectrum where the two peaks from $^{10}$B (left) and $^{10}$C (right) beams are clearly separated. 

\subsection{Beam Intensity.}
The total intensity of the incident beam ($^{10}$B+$^{10}$C) was measured using the ionization chamber. The ratio of the $^{10}$C to $^{10}$B particles in the beam was determined from the energy-loss spectrum. The total beam intensity multiplied by this ratio and the live-time of the data acquisition (DAQ) therefore provided a measure of the incident $^{10}$C beam counts on the target. This was used for obtaining the elastic scattering cross section and was determined for each instant of time throughout the experiment. The live-time fraction of the DAQ was ~ 0.87. The tail of the $^{10}$B beam events in the $^{10}$C beam selection window was ~ 4\% which is included in the total uncertainty of the cross sections determined.

\end{document}